\input{aipcheck}

\documentclass[
    ,final            % use final for the camera ready runs
  ]
  {aipproc}

\layoutstyle{8x11single}

\begin{document} 

\title{$X_{max}^{\mu}$ vs. $N^{\mu}$ from Extensive Air Showers as estimator for the mass of primary UHECR's. Application for the Pierre Auger Observatory.}

\classification{96.50.sd, 98.70.Sa, 13.85.Tp}
\keywords      {UHECR's, Extensive Air Showers, Muon arrival times, Mass composition}

\author{Nicusor Arsene}{
  address={Institute of Space Science, P.O.Box MG-23, Ro 077125 Bucharest-Magurele, Romania}
  ,altaddress={Physics Department, University of Bucharest, Bucharest-Magurele, Romania}
}

\author{Octavian Sima}{
  address={Physics Department, University of Bucharest, Bucharest-Magurele, Romania}
}

\begin{abstract}
We study the possibility of primary mass estimation for Ultra High Energy Cosmic Rays (UHECR's) using the $X_{max}^{\mu}$
(the height where the number of muons produced on the core of Extensive Air Showers (EAS) is maximum) and the number $N^{\mu}$ of muons 
detected on ground. We use the 2D distribution - $X_{max}^{\mu}$ against $N^{\mu}$ in order to find its sensitivity to the mass of the primary 
particle. For that, we construct a 2D Probability Function $Prob(p,Fe \ | \  X_{max}^{\mu},N^{\mu})$ which estimates the probability 
that a certain point from the plane $(X_{max}^{\mu}$, $N^{\mu})$ corresponds to a shower induced by a proton, respectively an iron nucleus. 
To test the procedure, we analyze a set of simulated EAS induced by protons and iron nuclei at energies of $10^{19} eV$ and $20^{\circ}$ 
zenith angle with CORSIKA. Using the Bayesian approach and taking into account the geometry of the infill detectors from the Pierre Auger 
Observatory, we observe an improvement in the accuracy of the primary mass reconstruction in comparison with the results obtained 
using only the $X_{max}^{\mu}$ distributions.
\end{abstract}
\maketitle
%%%%%%%%%%%%%%%%%%%%%%%%%%%%%%%%%%%%%%%%%%%%%%%%%%%%%%%%%%%%%%%%%%%%%%%%%%%%%%%%%%%%%%%%%%%%%%%%%%%%%%%%%%%%%%%%%%%%%%%%%%%%%%%%%%%%%%%%
\section{Introduction}
Nowadays, many experiments try to answer the fundamental questions about the origin, acceleration mechanism, propagation, mass composition and 
energy spectrum of UHECR's. One of these experiments is the Pierre Auger Observatory, located in Southern hemisphere in Argentina and dedicated 
to measure the proprieties of EAS induced by particles with energies greater than $10^{18} eV$. In this work we focus on 
a method by which one can reconstruct the mass of the primary UHECR using the surface detectors (SD) from the Pierre Auger Observatory which have 
a duty cycle of about $100\%$. 

%%%%%%%%%%%%%%%%%%%%%%%%%%%%%%%%%%%%%%%%%%%%%%%%%%%%%%%5%%%%%%%%%%%%%%%%%%%%%%%%%%%%%%%%%%%%%%%%%%%%%%%%%%%%%%%%%%%%%%%%%%%5%%%%%%%%%%%%
\section{Muon arrival times from EAS}
\begin{figure}
  \includegraphics[height=.29\textheight]{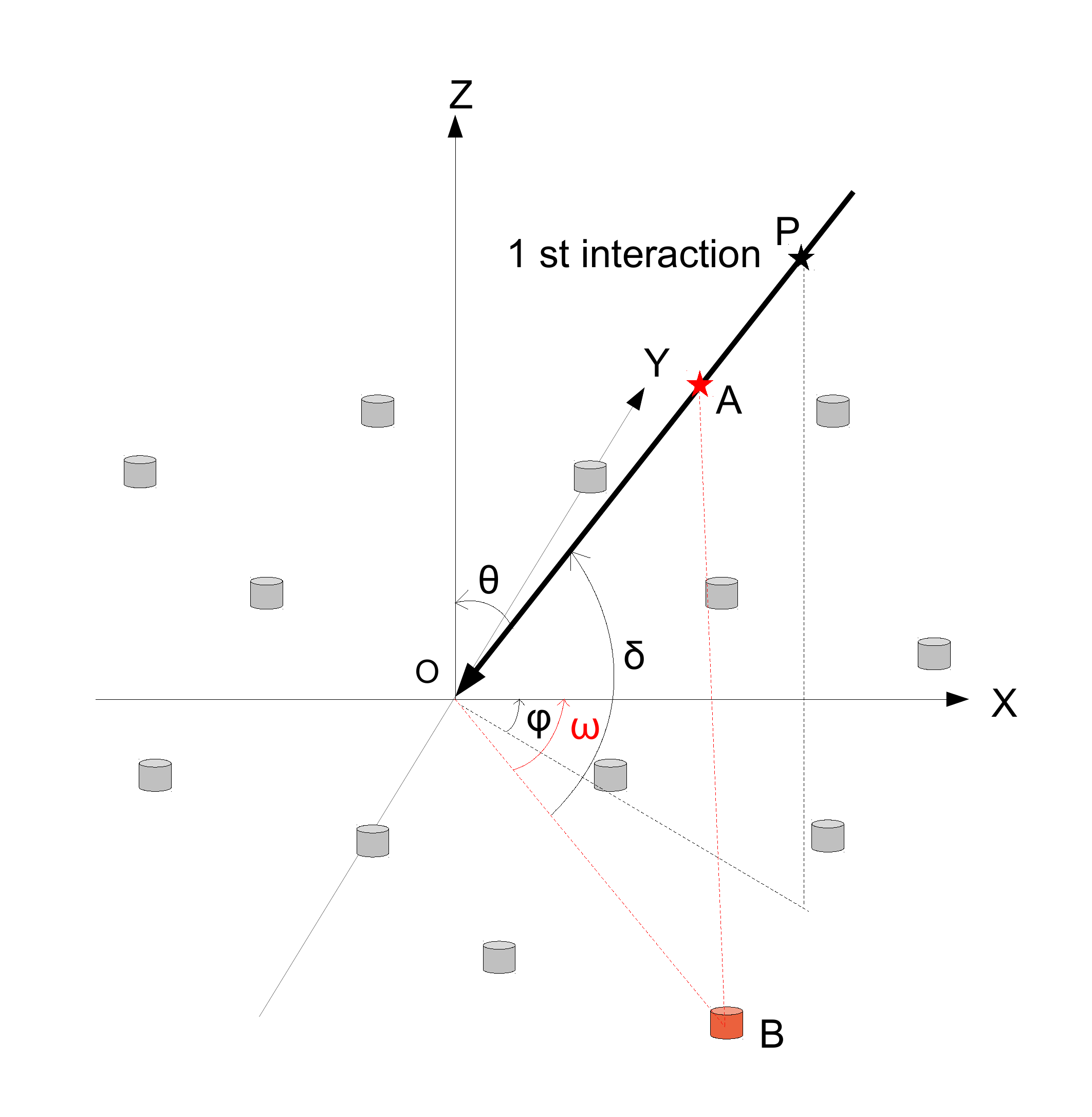}
  \caption{Coordinates system according with those from CORSIKA.}
  \label{coordinates_system1}
\end{figure}

This method has been previously proposed in \cite{Rebel:1994ed, Brancus, Haeusler, Cazon}.
The idea is to reconstruct the longitudinal profile of the muons produced on the shower core considering the times when the muons reach 
the ground relative to the shower core. Due to the larger cross section of the primary iron nuclei when compared to primary protons at the same energy,
it is obvious that the $<X_{max}^{\mu}>$ for protons is greater than the $<X_{max}^{\mu}>$ for iron induced showers. 
In Fig.~\ref{coordinates_system1} we have represented the coordinate system of the shower axis, which is the same as the one used in the CORSIKA code
\cite{corsika, corsika1}.We can calculate the time $t_{0}$ between the primary interaction ($P$) and the time when the muon is produced ($A$) \cite{Arsene}:
\begin{equation}
 {t_{0}} = \frac {(OB)^{2} - 2 c (OB)t_{c} \cdot cos\delta + c^{2}t_{c}^{2} - c^{2}t_{\mu}^{2}}  {2 c (c t_{c} - c t_{\mu} - (OB) cos\delta)}
\end{equation}
where $c$ represents the speed of light, $t_{c}$ and $t_{\mu}$ represent the time when the shower core respectively the muon reaches the ground.
Having the distribution of the heights at which all the muons were produced, we can transform it in units of $g/cm^{2}$ and then fiting with 
the Gaisser-Hillas function, the $X_{max}^{\mu}$ value can be obtained.
%%%%%%%%%%%%%%%%%%%%%%%%%%%%%%%%%%%%%%%%%%%%%%%%%%%%%%%%%%%%%%%%%%%%%%%%%%%%%%%%%%%%%%%%%%%%%%%%%%%%%%%%%%%%%%%%%%%%%%%%%%%%%%%%%%%%%%%%
\section{Results: $X_{max}^{\mu}$ vs. $N^{\mu}$ sensitivity to the primary mass}
\begin{figure}
  \includegraphics[height=.3\textheight]{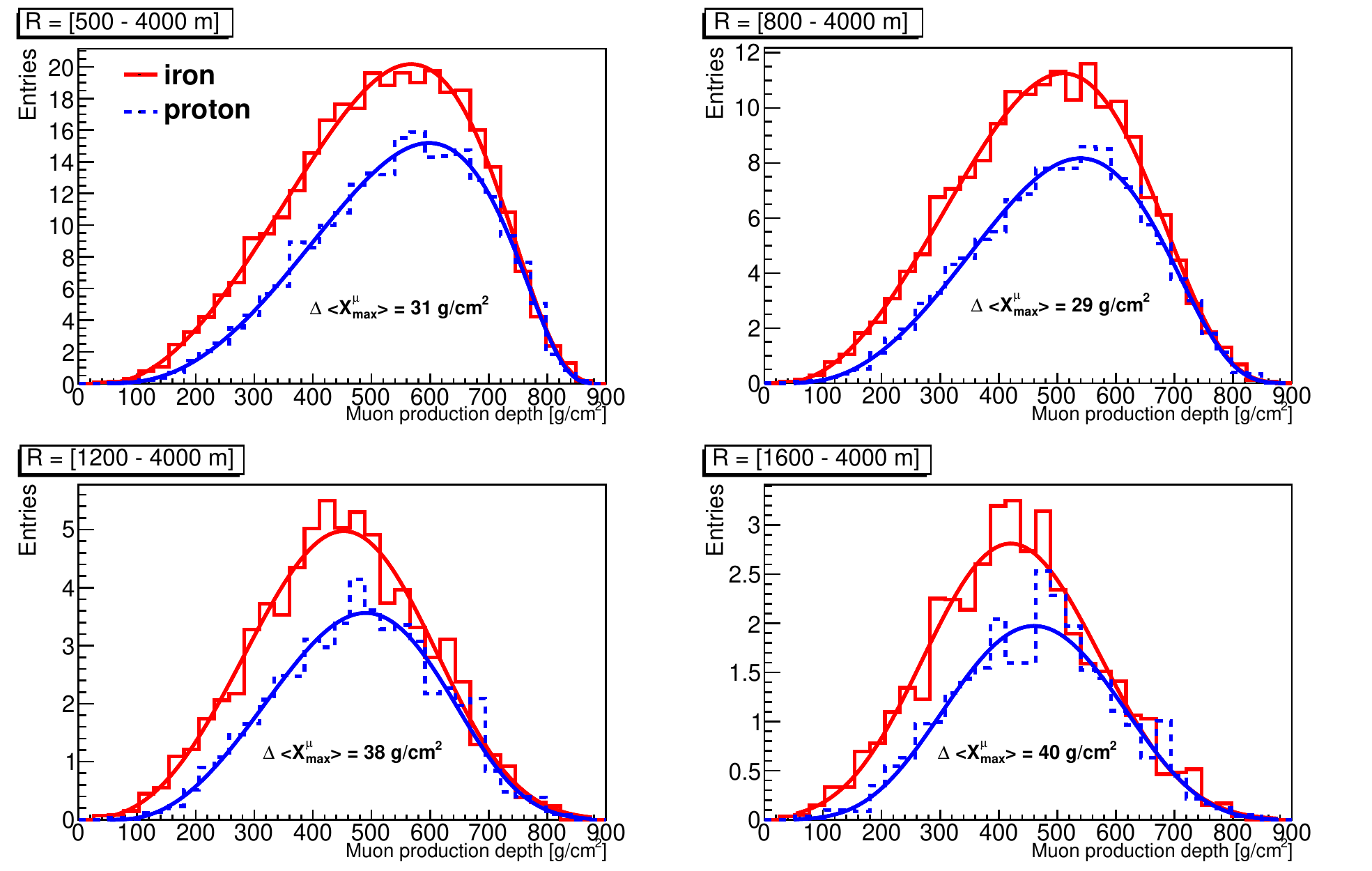}
  \caption{Longitudinal profile of muon production depth evaluated for different intervals of distances from shower axis. Average over 60 CORSIKA 
  simulations per case. We considered only the muons which hit the infill detectors of the Pierre Auger Observatory. The continuous 
  line represents the fit with the Gaisser-Hillas function.}
  \label{X_max_mu_vs_R}
\end{figure}

We performed a set of CORSIKA simulations containing 60 EAS induced by protons and 60 EAS induced by iron nuclei, at energies
$E = 10^{19} eV$, zenith angle $\theta = 20^{\circ}$, using the QGSJET-II \cite{Ostapchenko:2004ss} hadronic interaction model at the highest 
energies and taking into account the conditions from the Pierre Auger Observatory (observational plane, magnetic field, etc.). 
Using the method described above, Fig.~\ref{X_max_mu_vs_R} shows the reconstructed longitudinal profiles of the muons produced 
on the shower core considering different distances from the shower axis in observational plane.
\begin{figure}
  \includegraphics[height=.2\textheight ]{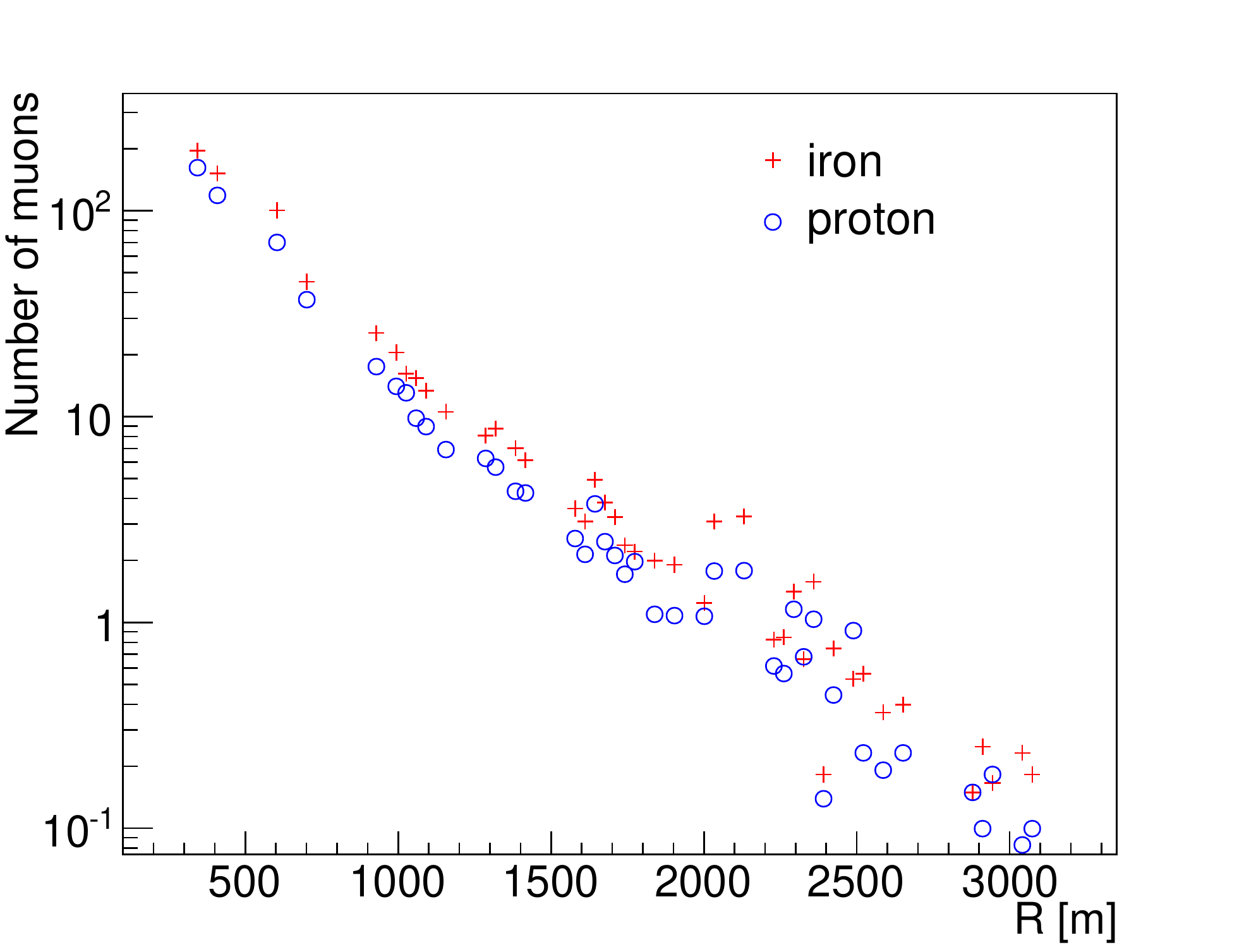}
  \caption{Number of muons produced on the shower core vs. the distance from the shower axis in the observational plane. Average over 60 
  CORSIKA simulations per case. We considered only the muons which hit the infill detectors of the Pierre Auger Observatory.}
  \label{Nrmu_vs_R}
\end{figure}
\begin{figure}
  \includegraphics[height=.5\textheight, angle=90, viewport=0cm 0cm 8cm 20cm,clip]{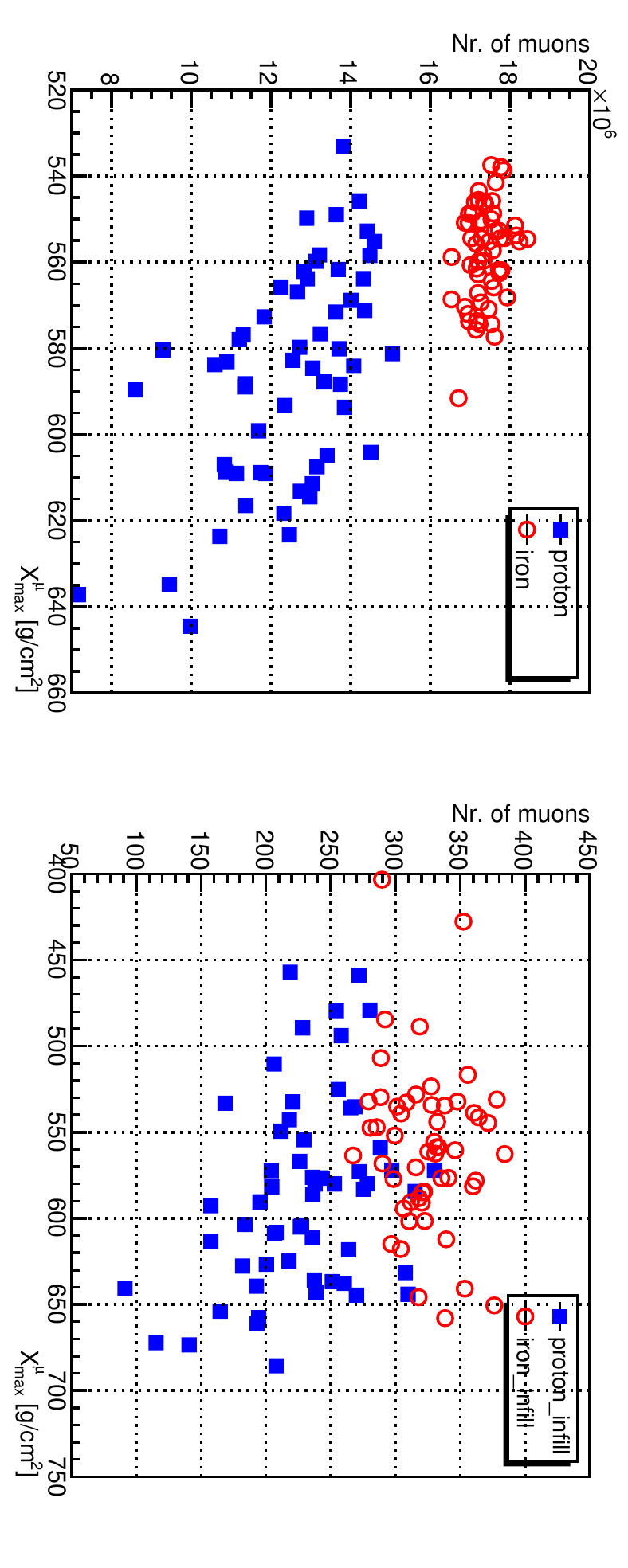}
  \caption{Event by event analysis, number of muons at ground level vs. $X_{max}^{\mu}$, 60 CORSIKA simulations per case. \textbf{Ideal} case, 
  when considering all the muons produced on the shower core at ground level \textit{(left)}, \textbf{infill} case when considering only the muons 
  which hit the infill detectors  \textit{(right)} }.
  \label{Nrmu_depth}
\end{figure}
It can be seen when using the muons from larger distances from the shower axis for the muon production depth (MPD) reconstruction, 
the difference between the maximum of the profiles tends to increase. In Fig.~\ref{Nrmu_vs_R}, the number of muons
produced on the shower core which reach the infill detectors is represented against the distance from the shower axis in observational plane for the protons, 
respectively iron nuclei induced showers in the same conditions. The larger number of muons on the ground for the case of iron nuclei as primary 
particle is due to the higher multiplicity at the first interaction when compared with the primary proton case. The dependence $X_{max}^{\mu}$ vs.
$N^{\mu}$ is represented in Fig.~\ref{Nrmu_depth}.
Having these two complementary informations $X_{max}^{\mu}$ and $N^{\mu}$ we define a 2D Probability Function
$Prob(p,Fe \ | \  X_{max}^{\mu},N^{\mu})$ which estimates the probability that a certain point from the plane
$(X_{max}^{\mu}$, $N^{\mu})$ corresponds to a shower induced by a proton, respectively an iron nucleus:
\begin{equation}
Prob(X_{max}^{\mu},N^{\mu} \ | \ p,Fe) = C \cdot exp \left(- \frac {(X_{max}^{\mu} - <X_{max}^{\mu}>)^{2}} {2 \sigma_{X_{max}^{\mu}}^{2}}\right) \cdot exp \left(- \frac {(N^{\mu} - <N^{\mu}>)^{2}} {2 \sigma_{N^{\mu}}^{2}}\right).
\label{2DProbFunc}
\end{equation}
The fit parameters of the function are listed in Table~\ref{tab}. Because the amplitude of the 
$Prob(X_{max}^{\mu},N^{\mu} \ | \ p,Fe)$ function depends on the ratio of the different species of nuclei in nature 
(in our particular case, the ratio of primary iron / protons), we use the Bayesian approach to test the reconstruction 
accuracy of the method.
\begin{table}
\begin{tabular}{lrrrrr}
\hline
  & \tablehead{1}{r}{b}{$C$}
  & \tablehead{1}{r}{b}{$<X_{max}^{\mu}>$}
  & \tablehead{1}{r}{b}{$\sigma_{X_{max}^{\mu}}$}
  & \tablehead{1}{r}{b}{$<N^{\mu}>$} 
  & \tablehead{1}{r}{b}{$\sigma_{N^{\mu}}$} \\
\hline
\bf{p, ideal} &$5.53 \pm 1.07$ & $572.7 \pm 18.07$ & $50 \pm 22.6$ & $1.22 \pm 0.3 \times 10^7$ & $1.66 \pm 0.33 \times 10^6$\\
\bf{Fe, ideal} & $35.34 \pm 6.23$ & $557.1 \pm 2.06$ & $13.75 \pm 1.47$ & $1.73 \pm 0.01 \times 10^7$ & $5.48 \pm 0.04 \times 10^5$\\
\bf{p, infill} &$3.32 \pm 0.7$ & $622.9 \pm 14.02$ & $35 \pm 2.74$ & $221.5 \pm 10.15$ & $35 \pm 3.34$ \\
\bf{Fe, infill} & $4.47 \pm 1.34$ & $569.4 \pm 14.78$ & $25 \pm 1$ & $321.5 \pm 6.14$ & $28.01 \pm 17.06$ \\

\hline
\end{tabular}
\caption{ Parameter values obtained after the fit of the distributions $X_{max}^{\mu} \ vs. \ N^{\mu}$ from Fig.~\ref{Nrmu_depth} with 
the 2D Probability Function Eq.~\ref{2DProbFunc}.}
\label{tab}
\end{table}

\section{Bayesian approach to test the procedure}
To test this procedure we need to define certain \textit{Prior} probabilities and then calculate the \textit{Posterior} probabilities 
that a certain point from the plane $(X_{max}^{\mu}$, $N^{\mu})$ corresponds to a shower induced by a proton or an iron nucleus. We know from 
the simulations $Prob(X_{max}^{\mu},N^{\mu} \ | \ p,Fe)$ (the probability to have the point with the coordinates $X_{max}^{\mu}$ and $N^{\mu}$ if 
the primary particle was proton or an iron nucleus). Supposing certain \textit{Prior} probabilities $Prob_{i}(p)$ and $Prob_{i}(Fe)$ which 
represents the ratio abundance of the primary protons and iron nuclei, we can calculate the \textit{Posterior} probabilities:

\begin{equation}
 Prob_{a}(p \ | \ X_{max}^{\mu}, N^{\mu}) = K \cdot Prob(X_{max}^{\mu}, N^{\mu} \ | \ p) \cdot Prob_{i}(p) , 
\end{equation}
\begin{equation}
 Prob_{a}(Fe \ | \ X_{max}^{\mu}, N^{\mu}) = K \cdot Prob(X_{max}^{\mu}, N^{\mu} \ | \ Fe) \cdot Prob_{i}(Fe) , 
\end{equation}
where the constant $K$ can be calculated from the normalization: 
\begin{equation}
Prob_{a}(p \ | \ X_{max}^{\mu}, N^{\mu}) + Prob_{a}(Fe \ | \ X_{max}^{\mu}, N^{\mu}) = 1.
\end{equation}
\begin{figure}
  \includegraphics[height=.2\textheight]{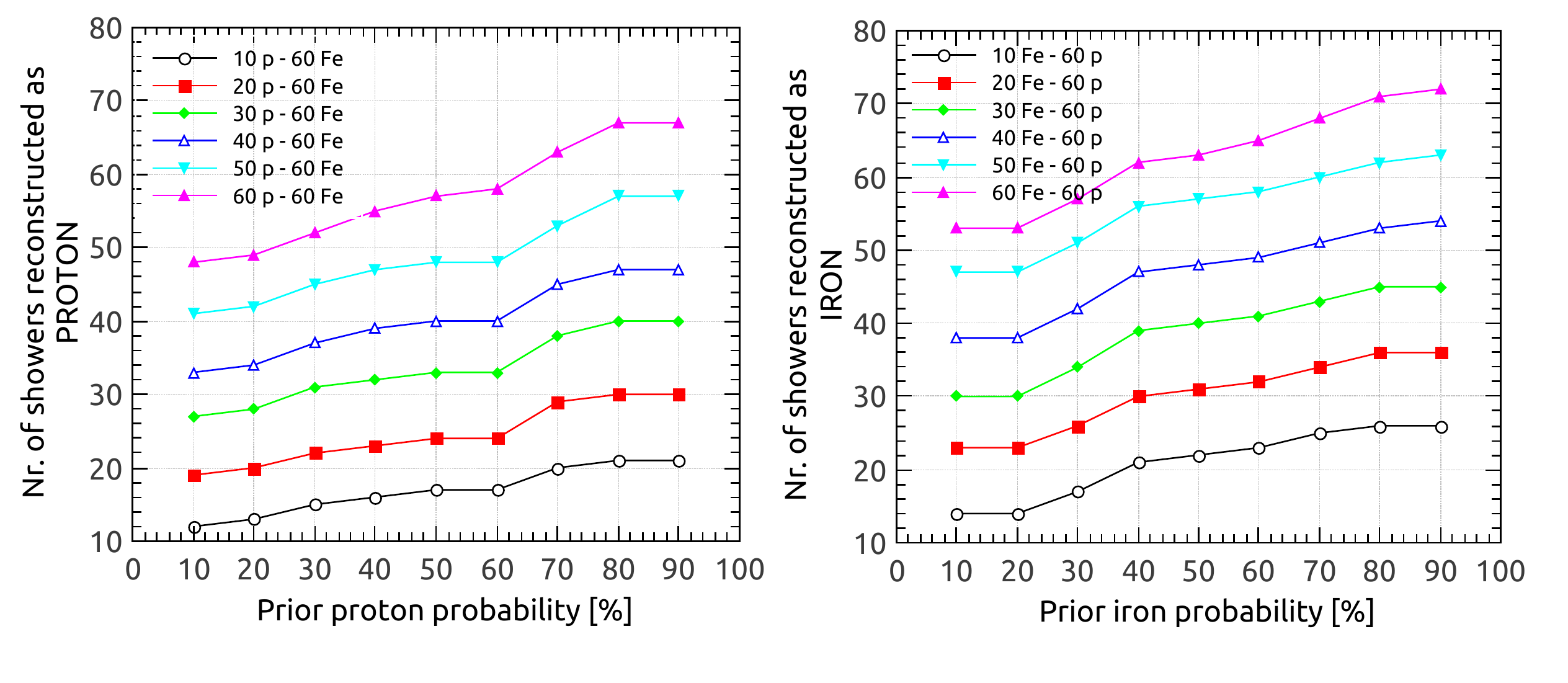}
  \caption{Potential mass discrimination of the method. Number of showers reconstructed as "PROTON" \textit{(left)} and "IRON" \textit{(right)} 
  for different prior probabilities and different mixtures of showers. Only the muons which entered in the infill detectors were taken into account.}
  \label{Bayes_prob_all}
\end{figure}
Fig.~\ref{Bayes_prob_all} shows how many showers are reconstructed to be induced by protons respectively iron nuclei, considering 
different prior probabilities and different combinations of number of showers.
In order to see the difference in the accuracy of the reconstruction methods, we consider the particular case with 60 proton induced showers 
and 60 iron induced showers, with the prior probabilities $Prob_{i}(p) = Prob_{i}(Fe) = 50\%$. Let's consider $Prob_{p \rightarrow p}$ to be the 
probability for proton induced showers to be reconstructed as "PROTON", respectively $Prob_{Fe \rightarrow Fe}$ the probability for iron 
induced showers to be reconstructed as "IRON". The comparison of the reconstruction accuracy of these two methods for this particular case can 
be seen in Table~\ref{tab1}.
\begin{table}
\begin{tabular}{lrr}
\hline
  & \tablehead{1}{r}{b}{$<X_{max}^{\mu}>$}
  & \tablehead{1}{r}{b}{($<X_{max}^{\mu}> vs. N^{\mu}$)} \\
\hline
\bf{$Prob_{p \rightarrow p}$} &$43.7 \pm 2.8 \%$ & $81.2 \pm 32.3 \%$ \\
\bf{$Prob_{Fe \rightarrow Fe}$} & $57.9 \pm 2 \%$ & $84.3 \pm 30.3 \%$ \\

\hline
\end{tabular}
\caption{Reconstruction accuracy of the methods $<X_{max}^{\mu}>$ and ($<X_{max}^{\mu}> vs. N^{\mu}$)}
\label{tab1}
\end{table}
\section{Conclusions}
We observed an improvement in the accuracy of the reconstruction of the primary UHECR's mass using the 2D Probability Function 
$Prob(X_{max}^{\mu},N^{\mu} \ | \ p,Fe)$ in comparison with the results obtained only with the $X_{max}^{\mu}$ distributions. 
Using this method with the surface detectors from the PAO, the duty cycle will be about $100\%$, and will 
considerably increase the statistics of the events with $E > 10^{19} eV$. This method will also be useful when searching for the 
quantum black holes signature proposed in \cite{Arsene:2013ria}.
\vspace{-5mm}
\begin{theacknowledgments}
N. Arsene thanks the organizers of the Carpathian Summer School for the fellowship. 
\end{theacknowledgments}
\vspace{-5mm}
\bibliography{sample}
\bibliographystyle{aipproc}

\IfFileExists{\jobname.bbl}{}
 {\typeout{}
  \typeout{******************************************}
  \typeout{** Please run "bibtex \jobname" to optain}
  \typeout{** the bibliography and then re-run LaTeX}
  \typeout{** twice to fix the references!}
  \typeout{******************************************}
  \typeout{}
 }

\end{document}